\documentstyle[prb,aps,twocolumn,epsfig]{revtex}
\begin{document}
\draft
\twocolumn[\hsize\textwidth\columnwidth\hsize\csname @twocolumnfalse\endcsname
\title{Systematic trends in the electronic structure parameters of 4$d$ transition
metal oxides Sr$M$O$_{3}$ ($M$ = Zr, Mo, Ru, and Rh)}
\author{Y. S. Lee}
\address{Center for Strongly Correlated Materials Research, Seoul National\\
University, Seoul 151-747, Korea}
\author{J. S. Lee and T. W. Noh}
\address{School of Physics and Research Center for Oxide Electronics, Seoul National\\
University, Seoul 151-747, Korea}
\author{Douck Young Byun and Kwang Soo Yoo}
\address{Department of Materials Science and Engineering, University of Seoul,\\
Seoul 130-743, Korea}
\author{K. Yamaura and E. Takayama-Muromachi}
\address{Superconducting Materials Center, National Institute for Materials Science,\\
1-1 Namiki, Tsukuba, Ibaraki 305-0044, Japan}
\date{\today }
\maketitle

\begin{abstract}
We investigated the electronic structures of the perovskite-type 4$d$
transition metal oxides Sr$M$O$_3$ ($M$ = Zr, Mo, Ru, and Rh) using their
optical conductivity spectra $\sigma (\omega )$. The interband transitions
in $\sigma (\omega )$ are assigned, and some important physical parameters,
such as on-site Coulomb repulsion energy $U$, charge transfer energy $\Delta
_{pd}$, and crystal field splitting $10Dq$, are estimated. It is observed
that $\Delta _{pd}$ and 10$Dq$ decrease systematically with the increase in
the atomic number of the 4$d$ transition metal. Compared to the case of 3$d$
transition metal oxides, the magnitudes of $\Delta _{pd}$ and 10$Dq$ are
larger, but those of $U$ are smaller. These behaviors can be explained by
the more extended nature of the orbitals in the 4$d$ transition metal oxides.
\end{abstract}


\vskip1pc] \newpage 

\section{Introduction}

There have been lots of investigations on 3$d$ transition metal oxides
(TMO), including cuprates and manganites, because they show a variety of
interesting electric and magnetic properties.\cite{cox92,Imada98} These
behaviors are closely related to the strong electron-electron ({\it el-el})
correlation, which originates from the localized 3$d$-orbitals. On the other
hand, 4$d$ TMO have attracted relatively less attention because it was
thought that the {\it el-el} correlation effects should be small and
insignificant due to their more extended $d$-orbitals. However, numerous
intriguing properties, such as superconductivity,\cite{Maeno95} non-Fermi
liquid behavior,\cite{Kostic98} pseudogap formation,\cite{yslee01} and
metal-insulator transitions,\cite{lee01,Katsufuji00,lee02} have been
observed recently in the 4$d$ TMO, especially ruthenates and molybdates.
These observations have stimulated new attention to the 4$d$ TMO.

Because 4$d$ TMO are characterized by more extended orbitals than 3$d$ TMO,
it has been generally believed that electrons in the extended 4$d$-orbitals
feel rather weak on-site Coulomb repulsion energy $U$ and exchange energy $J$%
, and the 4$d$-orbitals hybridize more strongly with neighboring orbitals,
e.g., O 2$p$-orbitals, than 3$d$-orbitals. Additional interactions, such as
spin-orbit coupling, also become significant.\cite{Mizokawa01} However,
these qualitative ideas are not sufficient to understand the intriguing
physical phenomena observed in some 4$d$ TMO. In 3$d$ TMO, systematic
investigations on $U$ and charge transfer energy $\Delta _{pd}$ provide a
basis to elucidate origins of numerous intriguing properties.\cite
{Imada98,Arima93,Bocquet96} Unfortunately, there have been few quantitative
studies about the electronic structures of 4$d$ TMO, except ruthenates,\cite
{lee01,Okamoto99} which makes it difficult to understand their physical
properties in more depth. Quantitative information on physical parameters
related to the electronic structures of TMO will serve as a starting
viewpoint in investigating various 4$d$ TMO with a potential to discover
other new intriguing phenomena. And, they will also allow us to make
comparisons with the 3$d$ TMO cases, which can provide us a better
understanding of physics of both TMO.

Optical spectroscopy is known to be a powerful tool to analyze the
electronic structures of TMO by probing the joint density of states between
unoccupied and occupied states. In this paper, we report a systematic
investigation on the electronic structures of the perovskite-type 4$d$ Sr$M$O%
$_3$ ($M$ = Zr, Mo, Ru, and Rh) by measuring their optical conductivity
spectra $\sigma (\omega )$. From these series with the same structure and
valency state $M^{4+}$, one can investigate how the electronic structures
change with $M$. As far as we know, our study is the first systematic effort
to investigate wide range optical properties of the 4$d$ Sr$M$O$_3$ series.
Based on proper electronic structure diagrams, the interband transitions
observed in their $\sigma (\omega )$ are assigned properly. From this, we
estimate important physical parameters, such as $\Delta _{pd}$, $U$, and
crystal field splitting energy 10$Dq$, which show systematic trends with $M$%
. Compared with the 3$d$ cases, the magnitudes of $\Delta _{pd}$ and 10$Dq$
are larger and those of $U$ are smaller in the 4$d$ Sr$M$O$_3$ compounds.
These behaviors can be understood as the more extended character of the 4$d$%
-orbitals than the 3$d$-ones. From our observations, it is found that these 4%
$d$ oxides belong to the Mott-Hubbard regime.

\section{Experimental techniques}

Polycrystalline SrZrO$_3$, SrMoO$_3$, and SrRhO$_3$ were prepared using the
solid state reaction technique. For the SrZrO$_3$ sample, SrCO$_3$ and ZrO$%
_2 $ were used as raw materials. After calcining and grinding repeatedly,
the resultant powders were pressed into a pellet under 200 MPa using cold
isostatic pressing. The SrZrO$_3$ pellet was sintered at 1700 $^{\text{o}}$C
for 5 hours. For the SrMoO$_3$ sample, the fine and pure SrO$_2$, MoO$_3$,
and Mo powders were mixed with a composition of SrMoO$_3$. The mixture of
approximately 0.2 g was placed into a gold capsule and then compressed at 6
GPa in a high-pressure apparatus. The sample was heated at 1300 $^{\text{o}}$%
C for 1 hour and quenched to room temperature at the elevated pressure. For
the SrRhO$_3$ sample, a preparation procedure similar to that of the SrMoO$%
_3 $ sample was used. The details of this procedure were published elsewhere.%
\cite{Yamaura01} The high pressure of sintering technique is effective to
provide metastable Mo$^{4+}$ and Rh$^{4+}$ states.\cite{Yamaoka92} From
x-ray diffraction measurements, it was confirmed that all the samples have a
single phase. From {\it dc} resistivity and magnetization measurements, it
was also found that their electric and magnetic properties are consistent
with the previous reports.\cite{Yamaura01,Agarwal99}

Just before optical measurements, we polished the sample surfaces up to 0.3 $%
\mu $m. Then, we measured reflectivity spectra from 5 meV to 30 eV at room
temperature. In the energy region between 5 meV and 0.6 eV, we used a
conventional Fourier transform spectrophotometer. Between 0.5 eV and 6.0 eV,
we used a grating spectrophotometer. And, in the deep ultraviolet region
above 6.0 eV, we used synchrotron radiation from the normal incidence
monochromator beam line at Pohang Accelerator Laboratory. After the optical
measurements, thin gold films were evaporated onto the samples and used for
making corrections for light scattering loss from the rough sample surfaces.%
\cite{lee99}

In order to obtain $\sigma (\omega )$ from the measured reflectivity
spectra, we performed the Kramers-Kronig (KK) analysis. It was known that
the KK analysis for the highly anisotropic polycrystalline samples could
provide incorrect $\sigma (\omega )$.\cite{orenstein88} However, all of our 4%
$d$ TMO have the slightly distorted perovskite structure and their optical
constants should be nearly isotropic, so the KK\ analysis could be applied
without any problem. For the analysis, the reflectivity below 5 meV was
extrapolated with the Hagen-Rubens relation for the metallic samples and
with a constant value for the insulating SrZrO$_3$ sample. For a high
frequency region, the reflectivity value at 30 eV was used for
reflectivities up to 40 eV, above which $\omega ^{-4}$ dependence was
assumed.\ To check the validity of our KK analysis, we independently
determined $\sigma (\omega )$ using spectroscopic ellipsometry techniques in
the photon energy range of 1.5 - 5.5 eV. The $\sigma (\omega )$ data from
the spectroscopic ellipsometry agreed quite well with the results from the
KK analysis.\cite{lee99}

\begin{figure}[tbp]
\epsfig{file=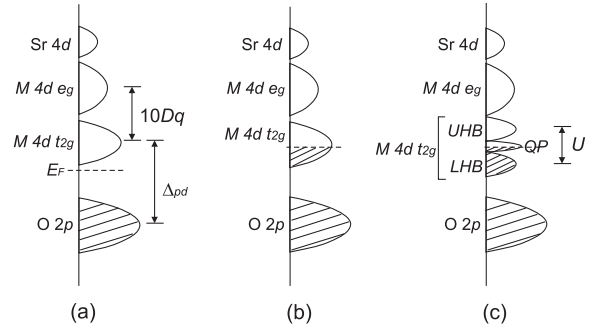,width=3.0in,clip=}
\vspace{2mm}
\caption{Schematic diagrams of the electronic structures of perovskite-type
4d Sr$M$O$_3$ with $M$ = Zr, Mo, Ru, and Rh. (a) SrZrO$_3$ (a band
insulator), (b) SrMoO$_3$ (a band metal), and (c) SrRuO$_3$ and SrRhO$_3$
(correlated metals). The dotted lines represent the Fermi level ($E_F$). In
(c), the partially-filled $t_{2g}$-band is split into occupied lower Hubbard
band (LHB) and unoccupied upper Hubbard band (UHB) by $U$ in addition to the
quasiparticle band (QP) located at $E_F$.}
\label{Fig:1}
\end{figure}
\section{Results and Discussions}

\subsection{Schematic diagrams of the electronic structures}

Figure 1 shows schematic diagrams of the electronic structures of the
perovskite-type Sr$M$O$_3$ compounds, where the 4$d$ transition metal $M$ is
either Zr, Mo, Ru, or Rh.\cite{footnote3} Figure 1(a) shows a typical
diagram of electronic structure for a $d^0$-insulator. SrZrO$_3$ is known to
be a 4$d^0$-insulator with a bandgap between the O 2$p$- and the Zr 4$d$ $%
t_{2g}$-band.\cite{cox92} The $e_g$-band has a higher energy level by
10$Dq$ than the $t_{2g}$-band. Usually, the
Sr 4$d$-band is in a higher energy level than $M$ 4$d$-bands. \cite
{Okamoto99} As the atomic number of $M$ increases from Zr, extra 4$d$
electrons start to fill the $t_{2g}$-band partially without any significant
change in the overall feature of the band structure. Here, the details of
the partially-filled $t_{2g}$ orbitals can vary according to the {\it el-el}
correlation. When the {\it el-el} correlation is quite weak, i.e., 4$d$
bandwidth $W\gg $ $U$, the partially-filled band induces a band metallic
state, as shown in Fig. 1(b). Since SrMoO$_3$ is a 4$d^2$-system\ and known
to be a Pauli paramagnetic band metal,\cite{Agarwal99} its electronic
structure should resemble Fig. 1(b). On the other hand, when the {\it el-el}
correlation is quite large, i.e., $U\gg $ $W$, the partially-filled band is
split into two Hubbard bands by $U$, which induces a Mott insulator. In the
intermediate state between two extreme cases, there can occur correlated
metals, whose band diagram is shown in Fig. 1(c). In the low-spin configuration,
the quasiparticle band is
located at $E_F$ in addition to the Hubbard bands, and the $e_g$-band
remains empty.\cite{lee01,Okamoto99} The electronic
structure of SrRuO$_3$, a low-spin 4$d^4$-system,\cite{Allen96} was
investigated by earlier workers,\cite{lee01,Okamoto99} and known to follow
Fig. 1(c). It was recently found that a 4$d^5$-system SrRhO$_3$ is a
paramagnetic correlated metal with the low-spin configuration.\cite
{Yamaura01} Because the $W$ of SrRhO$_3$ is narrower than that of SrRuO$_3$,
the Hubbard bands induced by the {\it el-el} correlation are more dominant
in SrRhO$_3$ than in SrRuO$_3$. So, the band diagram of SrRhO$_3$ can be
also explained by Fig. 1(c).

According to the Fermi golden rule,\cite{Fermi-golden} the {\it p-d}
transitions such as O 2$p$ $\rightarrow $ $M$ 4$d$ $t_{2g}$, $M$ 4$d$ $e_{g}$%
, and Sr 4$d$, should be distinct in $\sigma (\omega )$ of the 4$d$ Sr$M$O$%
_{3}$ systems. From such interband transitions, we can estimate some
physical parameters; a charge transfer energy $\Delta _{pd}$ from the O 2$p$ 
$\rightarrow $ $M$ 4$d$ $t_{2g}$ transition, and $10Dq$ from the energy
difference between O 2$p$ $\rightarrow $ $M$ 4$d$ $t_{2g}$ and $M$ 4$d$ $%
e_{g}$ transitions. On the other hand, the $U$ value should be estimated
from the {\it d-d} transition between two Hubbard bands, which should be
much weaker than the {\it p-d} transitions.

\subsection{Assignment of interband transitions in the Sr$M$O$_{3}$ compounds
}

Figure 2 shows $\sigma (\omega )$ of SrZrO$_3$ up to 20 eV. It is clearly
shown that this insulating $d^0$-compound has a large optical gap of $\sim $
5.6 eV, consistent with the previous reports.\cite{cox92} This value is
larger by $\sim $ 2 eV than the bandgap of SrTiO$_3$, i.e., 3.4 eV. Distinct
interband transitions are observed around 8 eV and 12 eV. To our knowledge,
there has been no band calculation report on this compound, so we assigned
these interband transitions by referring to the band structure of SrTiO$_3$,
which is a 3$d^0$ insulator. The dotted line in Fig. 2 represents $\sigma
(\omega )$ of SrTiO$_3$.\cite{palik91} Its overall featurea are very similar
to that of SrZrO$_3$, but with $\sim $ 2.5 eV shift to lower energy. K. van
Benthem {\it et al.} assigned the peaks around 5 eV and 9 eV of SrTiO$_3$\
as O 2$p$ $\rightarrow $ Ti 3$d$ $t_{2g}$ and O 2$p\rightarrow $ Ti 3$d$ $%
e_g $ transitions, respectively. \cite{Benthem01} [They also claimed that
the higher frequency peak should come from O 2$p\rightarrow $ Ti 3$d$ $e_g$
and/or Sr 4$d$.] Similarly, we can assign the peaks around 8 eV and 13 eV in
SrZrO$_3$ as O 2$p$ $\rightarrow $ Zr 4$d$ $t_{2g}$ and O 2$p\rightarrow $
Zr 4$d$ $e_g$ transitions, respectively. Note that these assignments are
consistent with the energy diagram of a band insulator, shown in Fig. 1(a).
By using the positions of the strong peaks, we can approximately estimate
that $\Delta _{pd}\sim 8$ eV and 10$Dq$ $\sim $ 5 eV in SrZrO$_3$, which are
larger than the values for SrTiO$_3$ (i.e., $\Delta _{pd}\sim $ 5 eV and 10$%
Dq\sim $ 4 eV).

\begin{figure}[tbp]
\epsfig{file=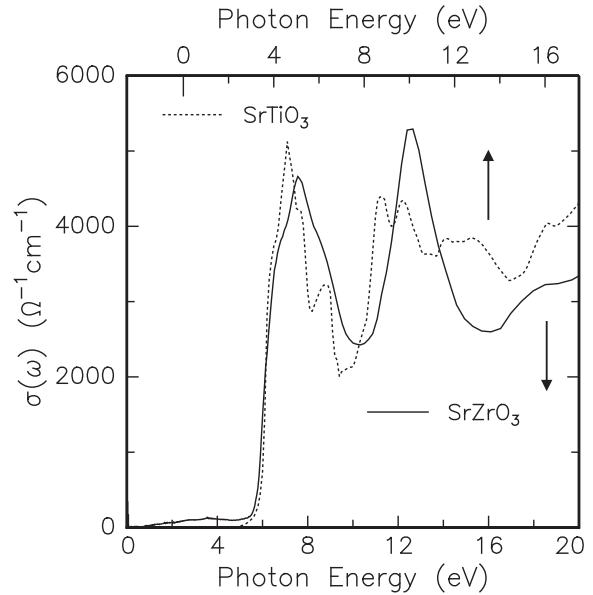,width=3.0in,clip=}
\vspace{2mm}
\caption{Room temperature $\sigma (\omega )$ of 4$d^0$ SrZrO$_3$ (the solid
line) up to 20 eV. The $\sigma (\omega )$ of $3d^0$ SrTiO$_3$ (the dotted
line), quoted from Ref. 21, is also displayed with a 2.5 eV shift to higher
energy for comparison with SrZrO$_3$. }
\label{Fig:2}
\end{figure}
Figure 3 shows $\sigma (\omega )$ of the Sr$M$O$_3$ series, with $M$ = Zr,
Mo, Ru, and Rh, up to 12 eV. The $\sigma (\omega )$ of SrRuO$_3$ is quoted
from our previous paper.\cite{lee01} The insulating SrZrO$_3$ has a
relatively large optical gap. For other metallic compounds, $\sigma (\omega
) $ below 1.0 eV have zero-frequency spectral weights, which decrease with
the increasing atomic number of $M$. The coherent mode of the band metallic
SrMoO$_3$ can be fitted well by the Drude model with a plasma frequency of $%
\sim $ 2.8 eV and a scattering rate of $\sim $ 0.3 eV. For other correlated
metallic SrRuO$_3$ and SrRhO$_3$, the low frequency $\sigma (\omega )$
decreases more slowly than the 1/$\omega ^2$-dependence which is predicted
by the Drude model. This behavior, which have often been observed in many
correlated metals,\cite{Imada98} indicates that the incoherent character in
the mid-infrared region might be rather strong.\cite{Kostic98}

The observed peaks of the Sr$M$O$_3$ compounds can be assigned according to
their electronic structures shown in Fig. 1. It is noted that, according to
the Fermi golden rule, the $p$-$d$ transition should be dominant in $\sigma
(\omega )$. The assignments for the SrZrO$_3$ peaks were already given. For
SrMoO$_3$, the 5.0 eV and 8.5 eV peaks are observed clearly, as shown in
Fig. 3(b). These two peaks can be assigned as O 2$p$ $\rightarrow $ Mo 4$d$ $%
t_{2g}$ and O 2$p\rightarrow $ Mo 4$d$ $e_g$ transitions. For SrRuO$_3$, the
3.0 eV, 6.0 eV, and 10 eV peaks shown in Fig. 3(c) can be assigned as O 2$p$ 
$\rightarrow $ Ru 4$d$ $t_{2g}$, O 2$p\rightarrow $ Ru 4$d$ $e_g$, and O 2$p$
$\rightarrow $ Sr 4$d$ transitions, respectively, according to our previous
report.\cite{lee01} Note that the {\it d-d} transition between the lower and
the upper Hubbard bands is located around 1.7 eV, but too weak to be clearly
seen in this compound.\cite{lee01} 
\begin{figure}[tbp]
\epsfig{file=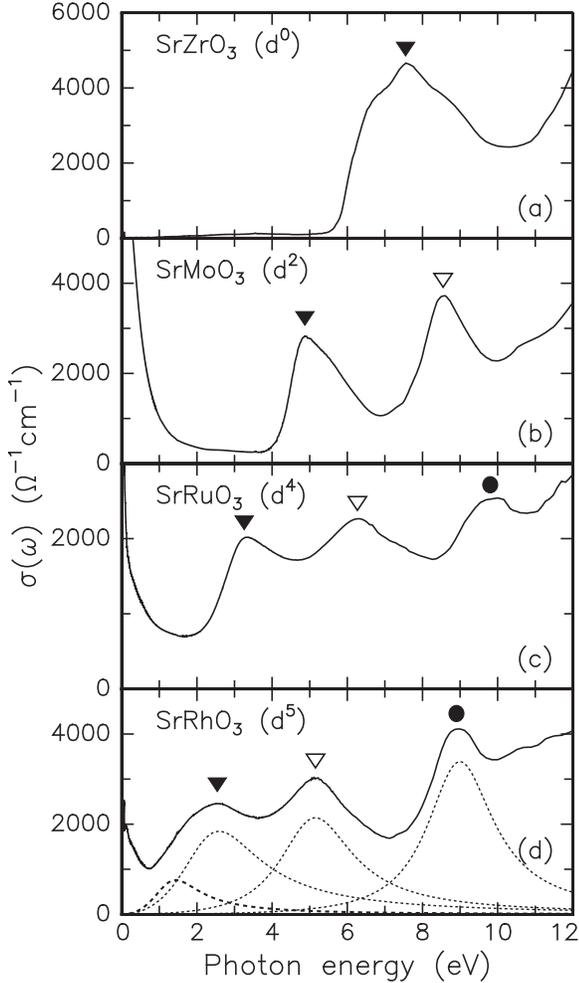,width=3.0in,clip=}
\vspace{2mm}
\caption{Room temperature $\sigma (\omega )$ of the 4d Sr$M$O$_3$ series.
(a) SrZrO$_3$ ($d^0$), (b) SrMoO$_3$ ($d^2$), (c) SrRuO$_3$ ($d^4$), and (d)
SrRhO$_3$ ($d^5$) up to 12 eV. The solid triangles, the open triangles, and
the solid circles represent the positions of the O 2$p$ $\rightarrow $ $M$ 4$%
d$ $t_{2g}$, the O 2$p\rightarrow $ $M$ 4$d$ $e_g$, and the O 2$p$ $%
\rightarrow $ Sr 4$d$ transitions, respectively. In (d), the dotted lines
represent the fitting results on the $\sigma (\omega )$ of SrRhO$_3$ with
Lorentz oscillators.}
\label{Fig:3}
\end{figure}

The interband transitions in SrRhO$_3$
can be similarly assigned for the case of SrRuO$_3$. However, the lowest
interband transition near 2.5 eV is quite asymmetric with a broad tail in
the low energy region, compared with the corresponding one in SrRuO$_3$.
From the electronic structure shown in Fig. 1(c), we fitted this peak with
two Lorentz oscillators, where a lower energy peak with a relatively small
strength can be assigned as the {\it d-d} transition between the Hubbard
bands. The fitting results are represented by the dotted lines in Fig. 3(d).
From this fitting, the 1.6 eV, 2.6 eV, 5.2 eV, and 9.0 eV peaks in SrRhO$_3$
can be assigned as the {\it d-d} transition between Hubbard bands, O 2$p$ $%
\rightarrow $ Rh 4$d$ $t_{2g}$, O 2$p\rightarrow $ Rh 4$d$ $e_g$, and O 2$p$ 
$\rightarrow $ Sr 4$d$ transitions, respectively. The relatively stronger 
{\it d-d} transition in SrRhO$_3$ indicates that the carriers in this
compound should be more correlated than those in SrRuO$_3$.

It is interesting to observe systematic trends in the interband transitions
of the Sr$M$O$_3$ series. As it goes from SrZrO$_3$ to SrRhO$_3$, all of the 
{\it p-d} transitions shift to lower energies. In the photon energy region
up to 12 eV, only the O 2$p$ $\rightarrow $ $M$ 4$d$ $t_{2g}$ transition is
observed in SrZrO$_3$, but the O 2$p$ $\rightarrow $ $M$ 4$d$ $e_g$
transition is additionally observed in SrMoO$_3$. And, in SrRuO$_3$ and SrRhO%
$_3$, the O 2$p$ $\rightarrow $ Sr 4$d$ transition as well as the O 2$p$ $%
\rightarrow $ $M$ 4$d$ transition are observed. Note that as the atomic
number of $M$ increases, the O 2$p$ $\rightarrow $ $M$ 4$d$ transitions
shift to the lower energy side. In addition, the peak interval between the O
2$p$ $\rightarrow $ $M$ 4$d$ $t_{2g}$ and the O 2$p$ $\rightarrow M$ 4$d$ $%
e_g$ transitions decreases. These interesting trends in peak positions
should be originated from the systematic changes of $\Delta _{pd}$ and $10Dq$%
.

\subsection{Systematic trends in $\Delta _{pd}$ and $10Dq$}

Figure 4(a) shows a systematic trend in the $\Delta _{pd}$ values, which are
estimated from the position of the O 2$p$ $\rightarrow $ $M$ 4$d$ $t_{2g}$
transition. As the atomic number increases, $\Delta _{pd}$ decreases.
According to J. B. Torrance {\it et al}.'s work with the ionic model,\cite
{torrance91} the change of $\Delta _{pd}$ with the atomic number of a
transition metal is attributed mainly to the change in electronegativity (or
ionization energy) of a transition metal; as the electronegativity of a
transition metal becomes larger, the $\Delta _{pd}$ decreases. So, the
decrease of $\Delta _{pd}$ in 4$d$ Sr$M$O$_3$ can be explained by the larger
electronegativity of $M$ with its atomic number increasing.

Figure 4(b) shows a systematic trend in the 10$Dq$ values, which are
estimated from the peak position difference between the O 2$p$ $\rightarrow $
$M$ 4$d$ $t_{2g}$ and the O 2$p$ $\rightarrow M$ 4$d$ $e_g$ transitions. The
10$Dq$ of SrZrO$_3$ is approximately estimated to be $\sim $ 5 eV, as shown
in Fig. 2. The $10Dq$ value of SrMoO$_3$, $\sim $ 3.8 eV, is comparable to
that in Mo 4$d$-bands of double perovskite Sr$_2$FeMoO$_6$, $\sim $ 4 eV,
from x-ray absorption spectroscopy.\cite{jung02} SrRhO$_3$ is estimated to
have 10$Dq$ $\sim $ 2.6 eV, a little smaller than that of SrRuO$_3$, i.e., 3
eV. It is noted that the $10Dq$ decreases with the increasing atomic number
of $M$. It is generally accepted that, as the overlap (or covalency) between
the O 2$p$- and the $d$-orbital becomes stronger, $10Dq$ becomes larger. So,
the decrease of the $10Dq$ in the 4$d$ Sr$M$O$_3$ series can be understood
as the shrinking of the $d$-orbitals, and the resultant weakening of the
covalency between the $M$ 4$d$- and the O 2$p$-orbitals with the increasing
atomic number. For a quantitative analysis, we estimate the covalency
strength as the {\it p-d} matrix element with the $\sigma $-bonding, $%
V_{pd\sigma }\propto d_r^{1.5}/d_{M-O}^{3.5}$, suggested by W. A. Harrison.%
\cite{Harrison80,footnote2} Here, $d_r$ and $d_{M-O}$ are the radial size of
the $d$-orbital\cite{Harrison80} and the distance between the $M$ and the O
ions, respectively. As shown in the inset of Fig. 4(b), the value of $%
d_r^{1.5}/d_{M-O}^{3.5}$ in the 4$d$ compounds decreases with increasing
atomic number of $M$, consistent with the decrease of $10Dq$. This suggests
that the {\it p-d} covalency should play a crucial role in determining the $%
10Dq$ value.\cite{Sugano63}

\begin{figure}[tbp]
\epsfig{file=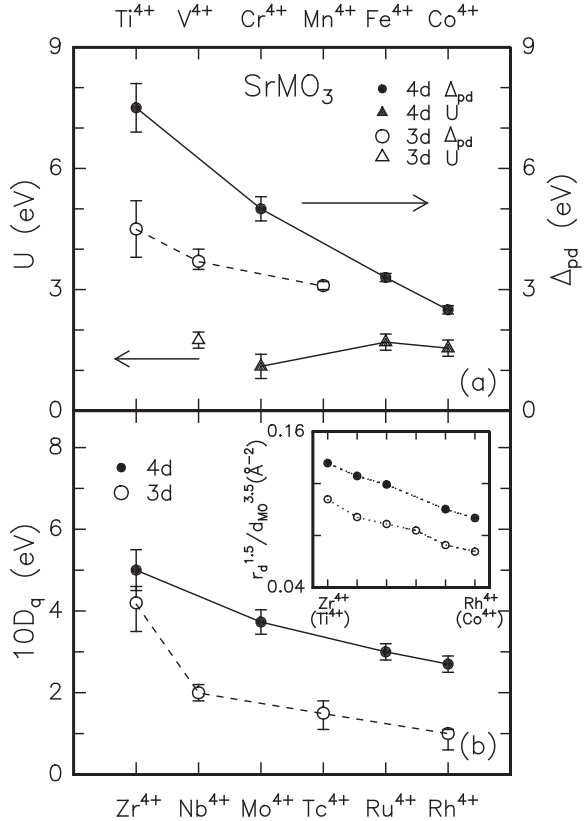,width=3.0in,clip=}
\vspace{2mm}
\caption{(a) Charge transfer energy $\Delta _{pd}$ and on-site Coulomb
repulsion energy $U$, and (b) crystal field splitting energy $10Dq$ in 4$d$
Sr$M$O$_3$ with $M$ = Zr, Mo, Ru, and Rh (solid symbols) and 3$d$ Sr$%
M^{\prime }$O$_3$ with $M^{\prime }$ = Ti,\protect\cite{Benthem01} V, 
\protect\cite{Vanadate} Mn,\protect\cite{Jung97,Manganite} and Co 
\protect\cite{Potze95} (open symbols). Note that these parameters are
estimated from the optical measurements only, except the $10Dq$ value of
SrCoO$_3$. \protect\cite{Potze95} In the inset of Fig. 4(b), the values of $%
d_r^{1.5}/d_{M-O}^{3.5}$ are estimated in the 3$d$ and the 4$d$ oxides. The
values of $d_r$ are used from Ref. 25. [We also estimated $d_r$ from the
ionic size of $M^{4+}$ ($M^{\prime 4+}$). The results show the similar
trends, except the case of SrFeO$_3$.] The $d_{M-O}$ of the 3$d$ compounds
are used from Ref. 23. For the 4$d$ oxides, because detailed structural
analyses have not been done, we used the values of $d_{M-O}$ as half of the
pseudo-cubic lattice constants, which are obtained from x-ray diffraction
measurements. The pseudo-cubic lattice constants $a$ of SrZrO$_3$, SrMoO$_3$%
, SrRuO$_3$, and SrRhO$_3$ are 4.11 \AA , 3.97 \AA , 3.94 \AA , and 3.92 \AA
, respectively. The $a$ value of 4.02 \AA\ is used for SrNbO$_3$ [H. Hannerz 
{\it et al}., J. Solid State Chem. {\bf 147}, 421 (1999)]. }
\label{Fig:4}
\end{figure}

\subsection{Comparison with 3$d$ transition metal oxides}

We also display the reported results on some 3$d$ Sr$M^{\prime }$O$_3$ with
a 3$d$ transition metal $M^{\prime }$= Ti, V, Mn, and Co in Fig. 4. Most of
the displayed results were determined from $\sigma (\omega )$ in the same
way as was adopted in this paper. Note that we include only the optical
results except for the 10$Dq$ value of SrCoO$_3$. In Fig. 4(a), the values
of $\Delta _{pd}$ in the 3$d$ Sr$M^{\prime }$O$_3$ series are displayed as
the open circles.\cite{Benthem01,Vanadate,Jung97} Similar to the case of the
4$d$ TMO, $\Delta _{pd}$ decreases as the atomic number of $M^{\prime }$
increases. [While there are no optical reports on metallic SrFeO$_3$ and
SrCoO$_3$, photoelectron spectroscopy (PES) results claimed that their $%
\Delta _{pd}$ values be nearly zero.\cite{Bocquet92,Potze95}] However, the
magnitudes of $\Delta _{pd}$ in 4$d$ Sr$M$O$_3$ are relatively larger than
those of 3$d$ Sr$M^{\prime }$O$_3$ with the same electron occupancy in $d$%
-orbitals. As mentioned earlier, the systematic change of $\Delta _{pd}$ is
attributed to the change in electronegativity. The larger $\Delta _{pd}$
values in the 4$d$ Sr$M$O$_3$ series than those in the 3$d$ Sr$M^{\prime }$O$%
_3$ series can be also explained by the smaller electronegativity for 4$d$
transition metals.\cite{footnote0}

In Fig. 4(b), the reported values of $10Dq$ in the 3$d$ Sr$M^{\prime }$O$_3$
series are plotted as the open circles.\cite
{Benthem01,Vanadate,Potze95,Manganite} The general trend of $10Dq$ in the 3$%
d $ series is also similar to that in the 4$d$ series. It is interesting to
see that the $10Dq$ of 4$d$ Sr$M$O$_3$ is larger than that of 3$d$ Sr$%
M^{\prime }$O$_3$. This behavior is consistent with the larger values of $%
d_r^{1.5}/d_{M-O}^{3.5}$ in 4$d$ Sr$M$O$_3$ than those in 3$d$ Sr$M^{\prime
} $O$_3$, shown in the inset. It is evident that the more extended 4$d$%
-orbitals induce stronger {\it p-d} covalency, which causes the $10Dq$
values of the 4$d$ oxides to be larger than those of the 3$d$ oxides.

While the high-spin configuration is more prevalent in 3$d$ TMO, the
low-spin configuration can be more easily found in 4$d$ Sr$M$O$_3$. This
behavior should be closely related to the relatively larger $10Dq$ values in
the 4$d$ Sr$M$O$_3$ series. The spin-configuration in TMO is determined
according to the relative magnitude of $10Dq$ and $J$: a high-spin
configuration for $10Dq<J$, and a low-spin configuration for $10Dq>J$. \ It
is known that some 3$d$ TMO, such as Mn- and Fe-oxides, have the high-spin
configuration. For example, the values of $10Dq$ and $J$ in Mn-oxides are
estimated to be 1.1 $\sim $ 1.8 eV\cite{Manganite} and $\sim $ 3 eV\cite
{Jung97}, respectively. On the other hand, 4$d$ TMO favor the low-spin
configuration with the relatively larger $10Dq$ and smaller $J$ due to the
more extended 4$d$-orbitals. For Ru-oxides, the values of $10Dq$ and $J$ are
estimated to be 3 eV and 0.5 $\sim $ 0.6 eV,\cite{Okamoto99,Singh96}
respectively. Although there is no report about $J$ in SrRhO$_3$, the large $%
10Dq$ value of $\sim $ 2.6 eV strongly suggests that this compound should
have the low-spin configuration. Because the Rh 4$d$-orbitals become more
extended in the Rh$^{3+}$ state than in the Rh$^{4+}$ state, an insulating 4$%
d^6$ LaRhO$_3$ material with a Ru$^{3+}$ state is very likely to be a
band-insulator which has fully-occupied $t_{2g}$ orbitals in the low-spin
configuration.

In the 3$d$ TMO, the character of the low energy charge excitations has
been investigated intensively.\cite{Imada98,Arima93,Bocquet96} According to
Zaanen, Sawatzky, and Allen's picture,\cite{ZSA85} the charge excitation
should be the $d$-$d$ transition between the Hubbard bands, if $U<\Delta
_{pd}$. This is classified as the Mott-Hubbard regime. If $U>\Delta _{pd}$,
the low energy excitation should be a $p$-$d$ transition, and this regime is
classified as the charge transfer regime. While the early 3$d$ TMO are
classified to be in the Mott-Hubbard regime ($U<\Delta _{pd}$), the late 3$d$
TMO mainly fall in the charge-transfer regime ($U>\Delta _{pd}$), implying
some kind of crossover between the two regimes.\cite{Imada98,Arima93} This
behavior is related to the increase of $U$ and the decrease of $\Delta _{pd}$
as the atomic number of a transition metal increases.

To find the character of the low energy charge excitation in the 4$d$ Sr$M$O$%
_3$ series, we estimated the values of $U$ from the position of the {\it d-d}
transition between the Hubbard bands in $\sigma (\omega )$. The values of $%
U $ for SrMoO$_3$ and SrRuO$_3$ are estimated from the previous studies. 
\cite{lee01,Taguchi02} Note that the $U$ value obtained from $\sigma
(\omega )$ is usually somewhat smaller than that measured from PES, due to
the exciton effects, but its difference between optical spectroscopy and PES
is usually less than 1.0 eV.

In Fig. 4(a), the values of $U$ for the 4$d$ Sr$M$O$_3$ series are plotted
as solid triangles. It appears that the magnitude of $U$ in 4$d$ Sr$M$O$_3$
is relatively smaller than that in 3$d$ Sr$M^{\prime }$O$_3$. As shown in
Fig. 4(a), the value of $U$ $\sim $ 2 eV in SrVO$_3$ ($d^2$)\cite{Vanadate}
is comparable to those of SrRuO$_3$ and SrRhO$_3$, 1.6 - 1.7 eV.
Generally, the $U$ value increases as the size of the $d$-orbitals
decreases. Because the $d$-orbital shrinks as the the atomic number
increases, the late 3$d$ Sr$M^{\prime }$O$_3$, such as SrFeO$_3$ ($d^4$) and
SrCoO$_3$ ($d^5$), are expected to have larger $U$ than the early 3$d$ SrVO$%
_3$, as has been confirmed by many PES studies. \cite{Imada98,Bocquet96}
From this, it is clear that the $U$ value of SrRuO$_3$ and SrRhO$_3$ should
be smaller than those of SrFeO$_3$ and SrCoO$_3$. \cite{footnote1} So, we
can say safely that 4$d$ Sr$M$O$_3$ have smaller $U$ than 3$d$ Sr$M^{\prime
} $O$_3$, which is quite natural due to the more extended nature of the 4$d$%
-orbitals.

The larger value of $\Delta _{pd}$ than $U$ indicates that all of the 4$d$ Sr%
$M$O$_3$ compounds investigated in this study should belong to the
Mott-Hubbard regime. From our studies on SrRuO$_3$ and SrRhO$_3$, the
magnitude of $\Delta _{pd}$, 2.6 - 3.0 eV, is larger than that of $U$,
1.6 - 1.7 eV, which indicates that even these compounds belong to the
Mott-Hubbard regime. This implies that these 4$d$ TMO can be used for
investigating the Mott-Hubbard transition with the {\it el-el} correlation.
Indeed, the importance of the {\it el-el} correlation has been observed in
ruthenates and molybdates,\cite{lee01,Katsufuji00} This was rather
surprising, since it was thought that the {\it el-el} correlation effects
should be insignificant in 4$d$ TMO due to the extended nature of the 4$d$%
-orbitals. And, due to the small value of $U$, the multiplicity of the 4$d$
orbitals might play important roles in physical properties of some 4$d$ TMO.%
\cite{park01} It is highly desirable to reinvestigate physical properties of
some 4$d$ TMO in view of the correlation effects.

\section{Summary}

We reported quantitative studies on physical parameters such as charge
transfer energy $\Delta _{pd}$, on-site Coulomb repulsion energy $U$, and
crystal field splitting energy 10$Dq$ of the perovskite-type of various 4$d$
Sr$M$O$_3$ ($M$ = Zr, Mo, Ru, and Rh) by optical conductivity analyses.
While the systematic changes of these parameters with the transition metal
are similar to those for the case of 3$d$ transition metal oxides, their
magnitudes are different; the $\Delta _{pd}$ and the 10$Dq$ values are
relatively larger, while the $U$ value is relatively smaller. These
behaviors are explained by the more extended character in 4$d$-orbitals than
in 3$d$-orbitals, which distinguishes the physical properties of the 4$d$
compounds from those of the 3$d$ ones. The relatively larger $10Dq$ is
closely related to the low-spin configuration in 4$d$ Sr$M$O$_3$. Due to the
relatively smaller $U$ and larger $\Delta _{pd}$, it is very likely that
most 4$d$ transition metal oxides lie in the Mott-Hubbard regime. Although
their $U$ values are relatively small, some intriguing physical phenomena
with the correlation effects could occur in 4$d$ transition metal oxides.

\acknowledgments
We would like to thank Jaejun Yu, S.-J. Oh, and S. D. Bu at SNU and D. Y.
Jung at SKKU for useful discussions. This work was supported by the Ministry
of Science and Technology through the Creative Research Initiative program,
and by KOSEF through the Center for Strongly Correlated Materials Research.
The experiments at PLS were supported by MOST and POSCO.


\begin{references}
\bibitem{footnote2}  The {\it p-d} matrix element with the $\sigma $-bonding 
$V_{pd\sigma }=\eta _{pd\sigma }\cdot d_r^{1.5}/d_{M(M^{\prime })-O}^{3.5}$,
and the {\it p-d} matrix element with the $\pi $-bonding $V_{pd\pi }=\eta
_{pd\pi }\cdot d_r^{1.5}/d_{M(M^{\prime })-O}^{3.5}$. Here, $\eta _{pd\sigma
}=(-2.17)\eta _{pd\pi }$,\cite{Harrison80} and represents the anisotropic
covalency strength of the O 2$p$ orbitals with the $t_{2g}$- and the $e_g$%
-orbitals. So, the quantity, $d_r^{1.5}/d_{M(M^{\prime })-O}^{3.5}$ can be a
parameter related to the 10$Dq$ between the $t_{2g}$- and the $e_g$-orbitals.
\bibitem{footnote0}  As the Madelung potential difference between a
transition metal and O sites, $\Delta V_M$ increases, so does $\Delta _{pd}$%
. Because the lattice constants of 4$d$ S$M$O$_3$ are larger than those of 3$%
d$ S$M^{\prime }$O$_3$, $\Delta V_M$ is expected to be smaller in 4$d$ S$M$O$%
_3$ than in 3$d$ S$M^{\prime }$O$_3$. So, the contribution of $\Delta V_M$
cannot explain the larger $\Delta _{pd}$ of 4$d$ S$M$O$_3$ than those of 3$d$
S$M^{\prime }$O$_3$.
\bibitem{footnote1}  From PES results, the $U$ value of SrRuO$_3$ is $\sim $
3 eV,\cite{Okamoto99} while that of SrFeO$_3$ is $\sim $ 7.8 eV.\cite
{Bocquet92}
\end{references}
\end{document}